\documentclass{article} 

\usepackage[utf8]{inputenc}

\usepackage{iclr2018_workshop,times}
\usepackage{url}
\usepackage{multirow}
\usepackage{lipsum}
\usepackage{amsthm}
\usepackage{mathtools}
\usepackage{xfrac}
\usepackage[export]{adjustbox}
\usepackage{longtable}
\usepackage{booktabs}
\usepackage{colortbl}
\usepackage{xcolor}
\usepackage{graphicx}
\usepackage{subcaption}

\usepackage{hyperref}
\hypersetup{
    colorlinks=true,
    urlcolor=[rgb]{0.188, 0.498, 0.886},
    linkcolor=[rgb]{0.188, 0.498, 0.886},    
    filecolor=magenta,      
    citecolor=[rgb]{0.188, 0.498, 0.886},
}

\usepackage{algorithm} 
\usepackage{program} 
\usepackage{algorithmic}
\usepackage{listings}
\usepackage{geometry}

\usepackage{caption,setspace}
\DeclareCaptionFormat{listing}{\rule{\dimexpr0.9\columnwidth+17pt\relax}{0.4pt}\par\vskip1pt#1#2#3}
\newcounter{code}



\DeclareMathVersion{sans}
\SetSymbolFont{operators}{sans}{OT1}{cmbr}{m}{n}
\SetSymbolFont{letters}  {sans}{OML}{cmbrm}{m}{it}
\SetSymbolFont{symbols}  {sans}{OMS}{cmbrs}{m}{n}

\lstnewenvironment{sflisting}[1][]
  {\lstset{#1}\mathversion{sans}}{}

\usepackage[normalem]{ulem}

\definecolor{grayalias}{HTML}{3F4444}

\definecolor{bluealias}{HTML}{307FE2}

\title{\textbf{Robotics CTF} (RCTF),
a playground for robot hacking}

\author{\textbf{Gorka Olalde Mendia,} 
   \textbf{Lander Usategui San Juan,}
    \textbf{Xabier Perez Bascaran,}  
   \textbf{Asier Bilbao Calvo,}\\
   \textbf{Alejandro Hernández Cordero,}
   \textbf{Irati Zamalloa Ugarte},
   \textbf{Aday Muñiz Rosas},
   \textbf{David Mayoral Vilches},\\
   \textbf{Unai Ayucar Carbajo}
   \textbf{Laura Alzola Kirschgens},
    \textbf{Víctor Mayoral Vilches},
    \textbf{Endika Gil-Uriarte} 
    \\
   Alias Robotics S.L. \\
   Vitoria-Gasteiz, Araba - Álava\\
   Spain \\
}

\usepackage{pdfpages}

\begin{document}
\maketitle

\vspace{-1em}
\begin{abstract}


Robots state of insecurity is onstage. There is an emerging concern about major robot vulnerabilities and their adverse consequences. However, there is still a considerable gap between robotics and cybersecurity domains. For the purpose of filling that gap, the present technical report presents the Robotics CTF (RCTF), an online playground to challenge robot security from any browser. We describe the architecture of the RCTF and provide 9 scenarios where hackers can challenge the security of different robotic setups. Our work empowers security researchers to a) reproduce virtual robotic scenarios locally and b) change the networking setup to mimic real robot targets. We advocate for hacker powered security in robotics and contribute by open sourcing our scenarios.

.



\end{abstract}



\section{Introduction}

The robotics landscape is rapidly evolving. Robots are spreading and will soon be everywhere. Systems traditionally employed in industry are being replaced by collaborative robots, and an increasing amount of professional and consumer robots are introduced in people’s daily activities. Following Personal Computers (PCs) and smartphones, robots are called to be the next technological revolution. Withal, robot cybersecurity is being largely underestimated, since safety cannot be granted without security \cite{RobotHaz}.

Over the last decade, the domains of security and cybersecurity have been substantially democratized, attracting individuals to many sub-areas within security assessment. According to recent technical reports summarizing hacker's activity in different sectors  \cite{hackerreport2017, hackerreport2018}, most security researchers are currently reporting vulnerabilities in websites (70.8\%) or mobile phones (smartphones, 5.6\%), and there is only a marginal contribution to emerging technologies such as Internet of the Things (IoT) devices (2.6\%). 

To date, only some pioneering offensive security studies \cite{RVSS, hackingbeforeskynet} have yet published relevant data about robotics' state-of-insecurity, but it seems to be an emerging field of research. We believe that the main reasons for this lag have been twofold. In a first aspect, robot security is a complex subject from a technological perspective which requires an interdisciplinary array of backgrounds, including security researchers, roboticists, software engineers and hardware engineers. In a second aspect, there are few guidelines or tools, and little formal documentation to assess robot security\cite{sedgewick2014framework, nistframework}. However, recent contributions have shed some light on the need of taking into account systematic security on robot deployments, \emph{inter alia} \cite{RSF} \cite{KHALID2018}. 

Furthermore, some of the components of modern robotics such as the Robot Operating System\cite{quigley2009ros} (ROS, and its second version ROS 2.0) have been developed as research platforms, and were purposefully developed without any security concerns. Some recent work demonstrated that robots powered by ROS are deployed revealing major vulnerabilities and flaws or simply left unprotected\cite{2018arXiv180803322D}. As the current state of ROS security is under question by the hacker and researcher community, there have been laudable but discrete efforts among the roboticists by adopting early security implementations. Through projects like SROS\cite{white2016sros} or Secure ROS, other available research works have dealt with hardening particular aspects of ROS \cite{finnicum2011building, mcclean2013preliminary, dieber2016application, lera2016cybersecurity, jeong2017study, dieber2017security, lera2017cybersecurity, martin2018quantitative, 2018arXiv180504101G}. But, overall, those have been poorly explored in the practice. We believe  that the robotics community and the ROS community could both greatly benefit from an integrative collaborative effort in an offensive security approach for robotics.

In an attempt to raise awareness around robot security, in this paper, we present the Robotics CTF (RCTF), an online playground  that invites white-hat-hackers to challenge robot security easily. The Robotics CTF is designed to be an online game, available 24/7, launchable through any web browser and designed to learn robot hacking step by step. In the following section we discuss the architecture of the RCTF and introduce some of the basic available scenarios.

The remainder of the paper is structured as follows. Section \ref{prior} discusses previous work and in particular, the Capture the Flag competitions for hackers. Section \ref{rctfsection} presents Alias Robotics' Robotics CTF and Section \ref{conclusion} presents paper outline and future remarks. 


\section{Previous work}
\label{prior}

In the field of cybersecurity, Capture The Flag (CTF) competitions are designed as the outdoor game and computer game counterparts (e.g. Unreal Tournament, Counter Strike, etc). A flag is hidden somewhere over an scenario and teams or individuals attempt to capture it, scoring points accordingly. The flags consist of secret data and the player needs to exploit weaknesses present in code or binaries to capture it.

Games start by opening the problem environment to the hacker or team of hackers. Most commonly, different strategies are suitable to reach the flag, which gives raise to develop different tactics in what is known as the "Jeopardy" game mode. Under some conditions, there are "defensive" and "offensive" endeavours within the game ("attack/defense"), a mix of both or even other scopes, such as "King of the Hill" in which the control of a vulnerable entity is disputed over different hackers participating. 

In the conventional CTF conception, the "flag" itself can be a virtual flag or an abstract goal. \emph{Id est}, the target can be perceivable or hidden, depending on the scenario and the scope of the game. Under some conditions, the "flag" can be unknown, so that the player needs to develop a particular awareness and sharpen her/his perception of the Capture The Flag environment looking for unusual objects, aspects or behaviours. Other creative designs of CTFs mix creative story lines to engage the hackers into a story oriented to solve an array of problems that are presented into a story gameplay. The overall objective of CTFs is to score over the opponent(s), by achieving completion of tasks faster, for a longer duration or more effectively. Table \ref{tabular:ctftypes} shows different historical use cases and objectives of Capture The Flag competitions.

\renewcommand{\arraystretch}{2}
\begin{table}
\caption{Capture The Flag (CTF) methodologies and types.}
\label{tabular:ctftypes}
\begin{tabular}{ |p{2cm}| p{3cm}|p{3cm}|p{6cm}|  }
 \hline
 \textbf{Platform (Site)} & \textbf{Type} & \textbf{Performer} & \textbf{Objective} \\
 \hline
 \multicolumn{4}{|c|}{\textsc{Non-security oriented CTFs}} \\
 \hline
 Outdoor game   & Attack/Defense  & Teams & Find and capture a given physical flag in an outdoor environment and Protect your teams' flag \\
 Unreal Tournament&   Attack/Defense  & Teams of players (Humans or bots)   & Grab other teams flag to your own base, without being killed on the way, and protect yours against members of the other team\\
 Counter Strike & Attack/Defense & Teams of players (Humans or bots) &  \emph{Idem}\\
 \hline
 \multicolumn{4}{|c|}{\textsc{Security Oriented CTFs}} \\
 \hline
 DEFCON's CTF &   Attack/Defense  & Teams & On custom services, attack other's, patch and protect your own with secret scoring system\\
 iCTF & Attack/Defense  & Teams   & Maintain a set of services so that they remain available and uncompromised throughout the contest and attack others'\\
 DARPA's CGC & System and development oriented  & Teams & Automated hacker systems are programed by teams to perform an array of security tasks\\
 iHacklabs (classic) & Jeopardy   & Hackers / Teams & Complete challenges in web, reverse engenieering, forensics and criptography\\
 Hacknet & Jeopardy & Individual & Following "Bit's" story, the hacker needs to solve problems on the go \\
 NetKotH & King of the Hill & Hackers/Teams & The contestants must gain access to a challenge machine which may contain multiple vulnerabilities. Once they are on the challenge machine, they must plant their team tag where the ScoreBot can see it. \\
 Hack the Box & Jeopardy & Hackers & Contains several dynamic challenges. Some of them simulate real world scenarios and some appertain to CTF style of challenges \\
 FACEBOOK (Educational CTF) & Jeopardy and King of the Hill (Supported) & Hacker/Teams & Design your own CTF environments and competition \\
 \hline
\end{tabular}
\end{table}

CTF "gaming" techniques are propitious for initiation of new security exercises, displaying of novel pentesting capabilities in the hacker community, challenging novel targets of environment or emphasizing upon a particular security aspect \cite{eagle2004capture}. CTFs can be useful for educating audiences \emph{ad hoc}  and helps develop novel security skills. In capture the flag environments hackers, from experts to beginners, can show their capabilities of lateral thinking competing against other hackers, getting credit accordingly in the ranking of the player community. These new competences enable white-hat or ethical hackers to learn and identify new vulnerabilities, pinpointing them, which allows the owner of the system become aware of those before a third party with malicious intentions finds and exploits them. 

\begin{figure}[t!]
    \centering
    \begin{subfigure}[b]{0.45\textwidth}
        \includegraphics[width=\textwidth]{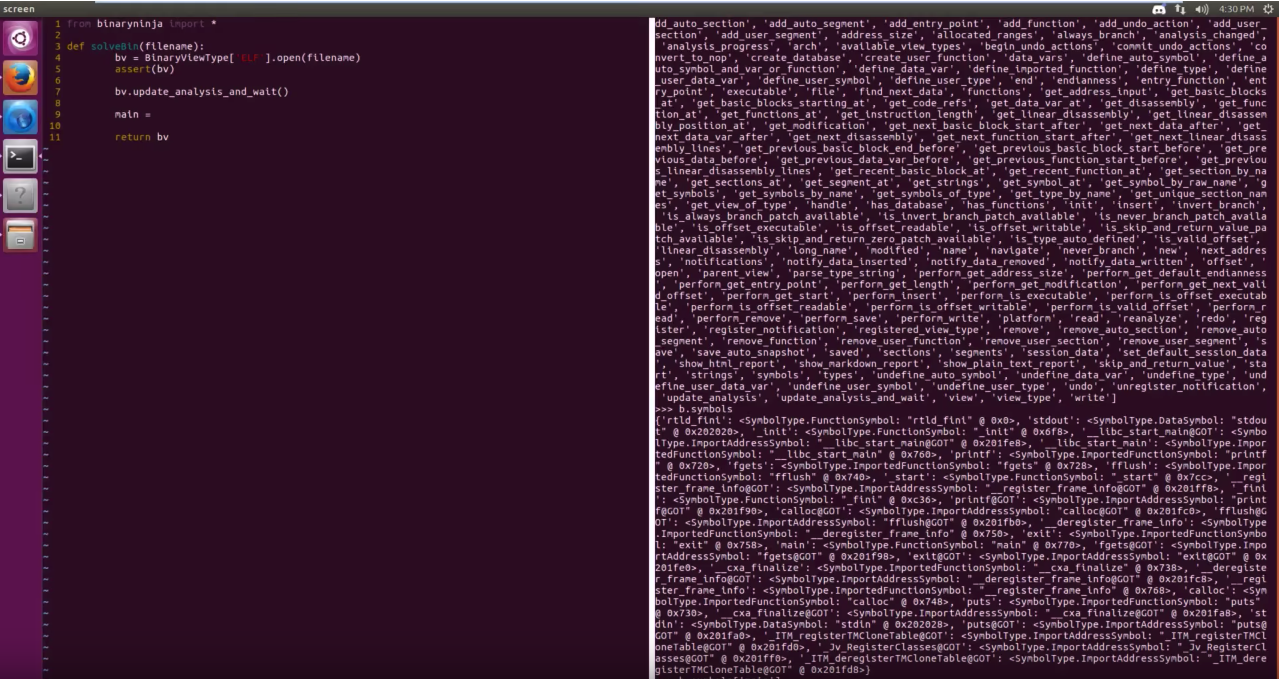}
        \caption{{DEFCON's CTF}}
        \label{fig:rctf1}
    \end{subfigure}
    ~ 
    \begin{subfigure}[b]{0.45\textwidth}
        \includegraphics[width=\textwidth]{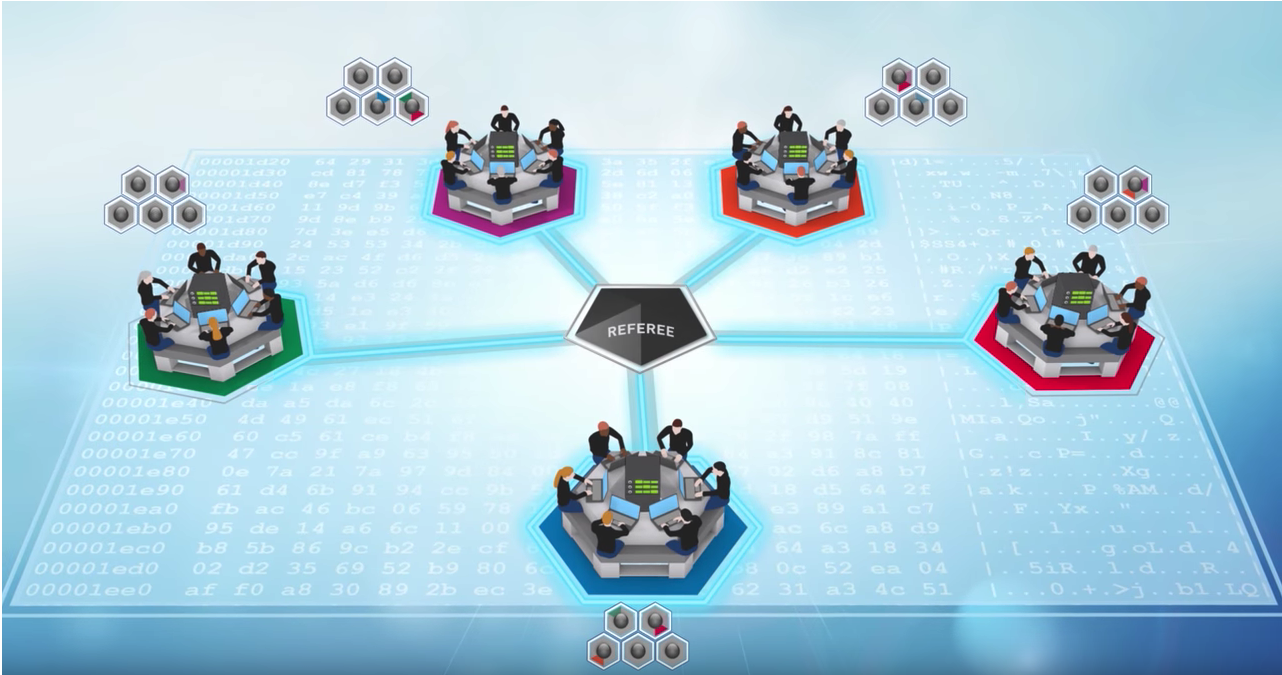}
        \caption{{DARPA's CGC}}
        \label{fig:rctf2}
    \end{subfigure}
    ~ 
    \begin{subfigure}[b]{0.45\textwidth}
        \includegraphics[width=\textwidth]{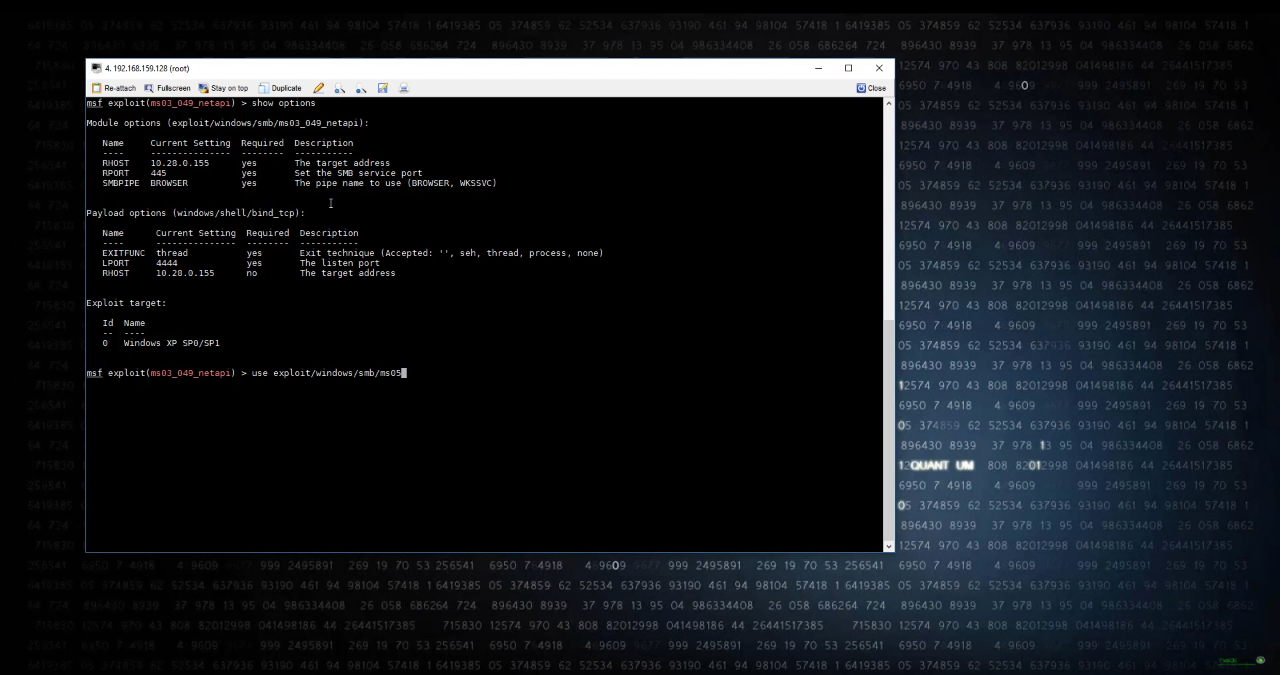}
        \caption{{iHackLabs CTF}}
        \label{fig:rctf3}
    \end{subfigure}
    ~
    \begin{subfigure}[b]{0.45\textwidth}
        \includegraphics[width=\textwidth]{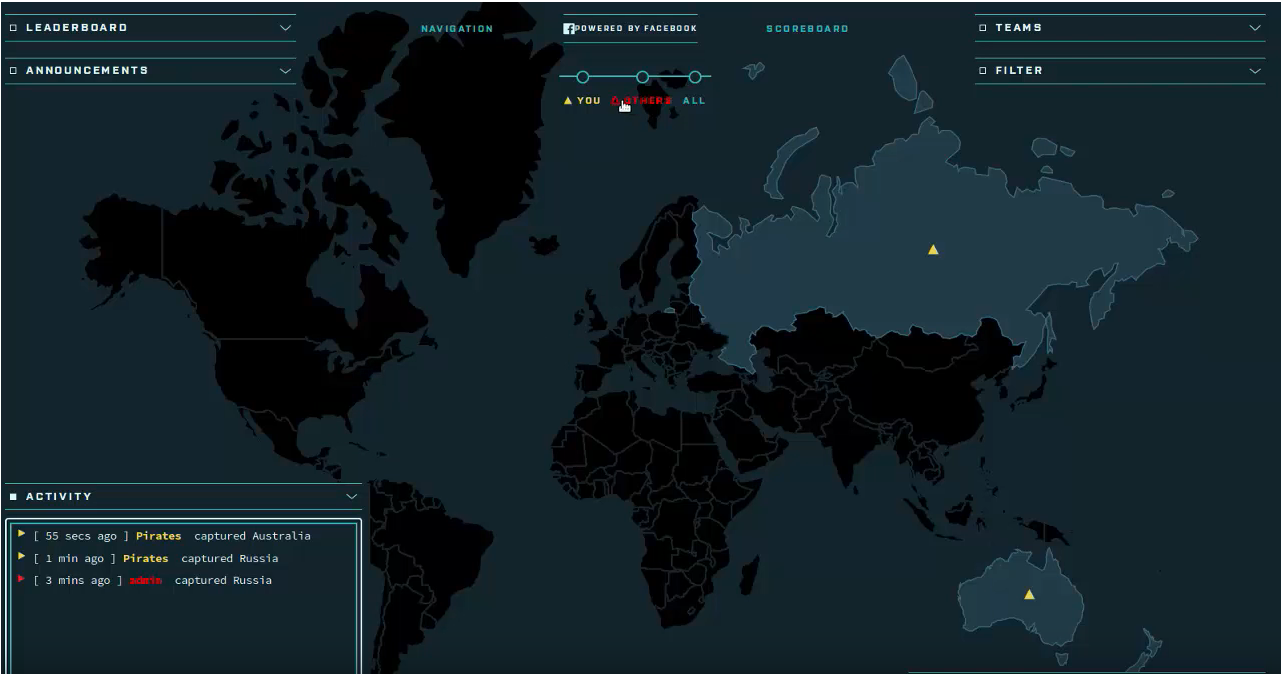}
        \caption{{FB CTF template model}}
        \label{fig:rctf4}
    \end{subfigure}

    \caption{{Snapshots of available Capture the Flag competitions}}
    \label{fig:rctf_peek}
\end{figure}

To date, CTF competitions have been designed in most facets of IT. There are some extremely well known CTFs such as DEFCON's \cite{cowan2003defcon}, the international capture the flag competition (iCTF) \cite{vigna20112010}, DARPA's Cyber Grand Challenge \cite{darpa2014cyber} and even dozens of successful companies devoted to create CTFs in the most creative fashions, which includes the most up to date aesthetics and state of the art front-end developments.  Figure \ref{fig:rctf_peek} illustrates real snapshots several relevant CTF competitions in the international panorama, each of which is adapted to a particular scenario with a unique security focus, as explained in Table \ref{tabular:ctftypes}.

Inspired by the state of the art, our research found out that, to the best of our knowledge, no CTF has focused or ever covered robotics as the leitmotif. Additionally, we believe that the CTF approach could be a very powerful and useful tool to help securing robots. We aim to engage security researchers and offensive security specialists into the largely unexplored robotics landscape.


\section{Robotics CTF (RCTF)}
\label{rctfsection}
Motivated by the growing insecurity in the field of robotics and the lack of security countermeasures adopted by robot manufacturers, our team is proud to introduce the first Robotics Capture The Flag game: the Robotics CTF (RCTF). The RCTF has been designed to be an online playground, available 24/7 and available through any browser from anywhere in the world, to learn robot hacking step by step. White-hat hacker audience is invited to test, challenge, learn and interact with state of the art of robot environments, from an educational perspective and with robot security as the final goal. Gradually, the accomplishments during the RCTF program enable the ethical hacker to acquire the competences to assess robot security.

To play the RCTF, a user needs to provide a valid e-mail address and accept the terms of use. In addition, each hacker  is kindly asked to behave decorously and to not act beyond the purpose of the gamification.

Alias Robotics' RCTF consists on an array of serial scenarios that hackers have to successfully complete as fast and accurately as possible, in order to proceed to the next scenario. With each completion, the successful robot hacker will be provided with a password that allows him/her to proceed to the next. The robot hacker can review her/his position on the ranking table and compare results against the rest of the hackers in the RCTF community.

\begin{figure}[h!]
\centering
 \includegraphics[width=0.7\textwidth]{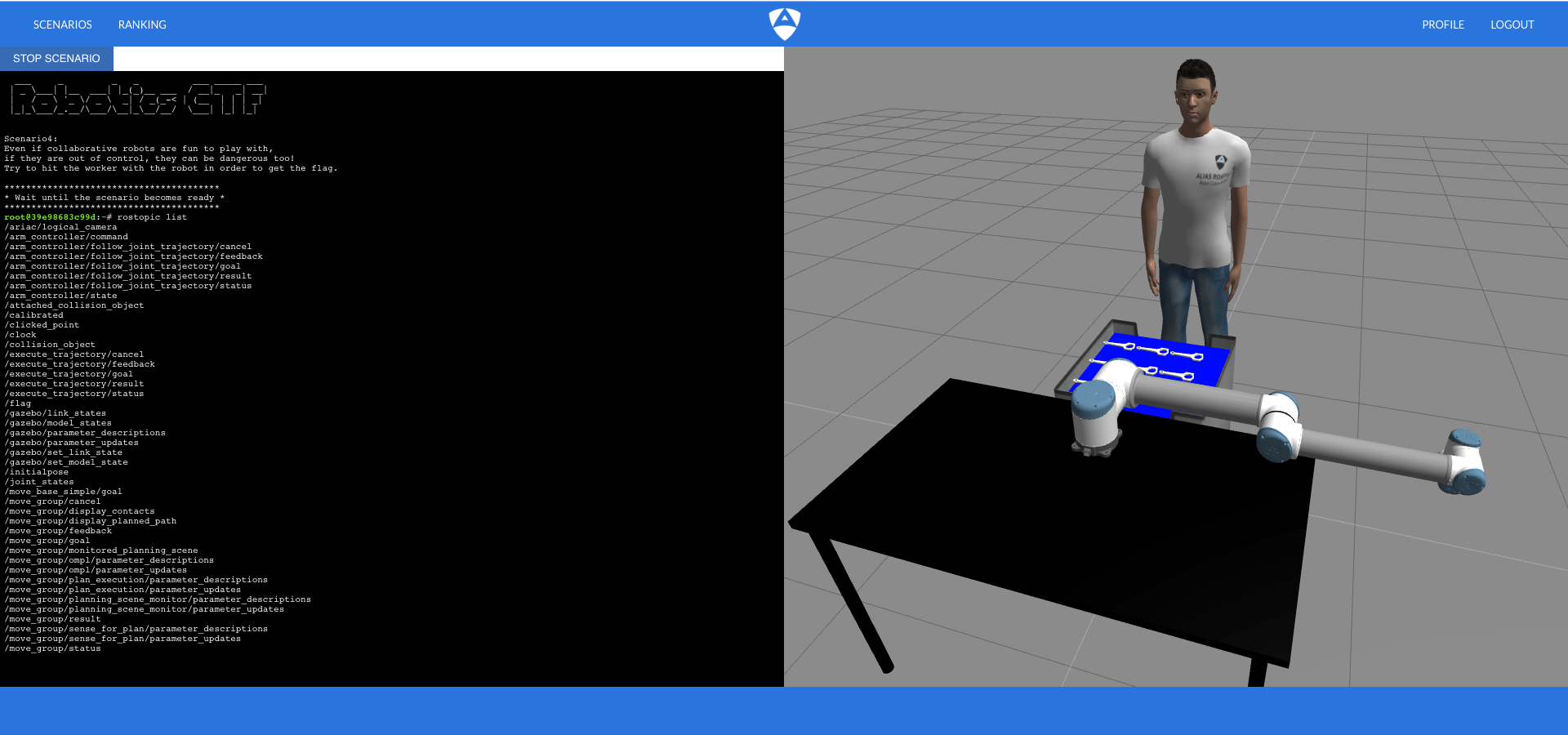}
\caption{\footnotesize Shows a snapshot of Robotics CTF (RCTF) Scenario 4 with a virtual collaborative robot (UR10) and Prudencio, wearing his Alias Robotics T-shirt. Imprudently, he does not realize that his UR10 is under the control of a third party and goes too close. The objective of this scenario is to hit "Pruden" with UR10 , which will unlock the flag and allow the robot hacker to proceed to the next level.}
\label{fig:scenario}
\end{figure}

Robotics CTF is designed to provide hackers with a full experience of the security landscape in robotics. Integrated in our webpage, the platform allows to learn using tools such as ROS, is compatible with other hacking tools and provides robot simulation through Gazebo \cite{Koenig04designand} as shown in Figure \ref{fig:scenario}. The first scenarios are education-oriented and, by achieving those, the hacker will gain basic know-how for the forecoming challenges. The scenarios depicted in RCTF (Table \ref{rctfscenarios}) are fictitious and do not have real-world counterparts, but do certainly reflect similarities with current real platforms in robotics.

\subsection{Architecture}

\begin{figure}[h!]
\centering
 \includegraphics[width=0.7\textwidth]{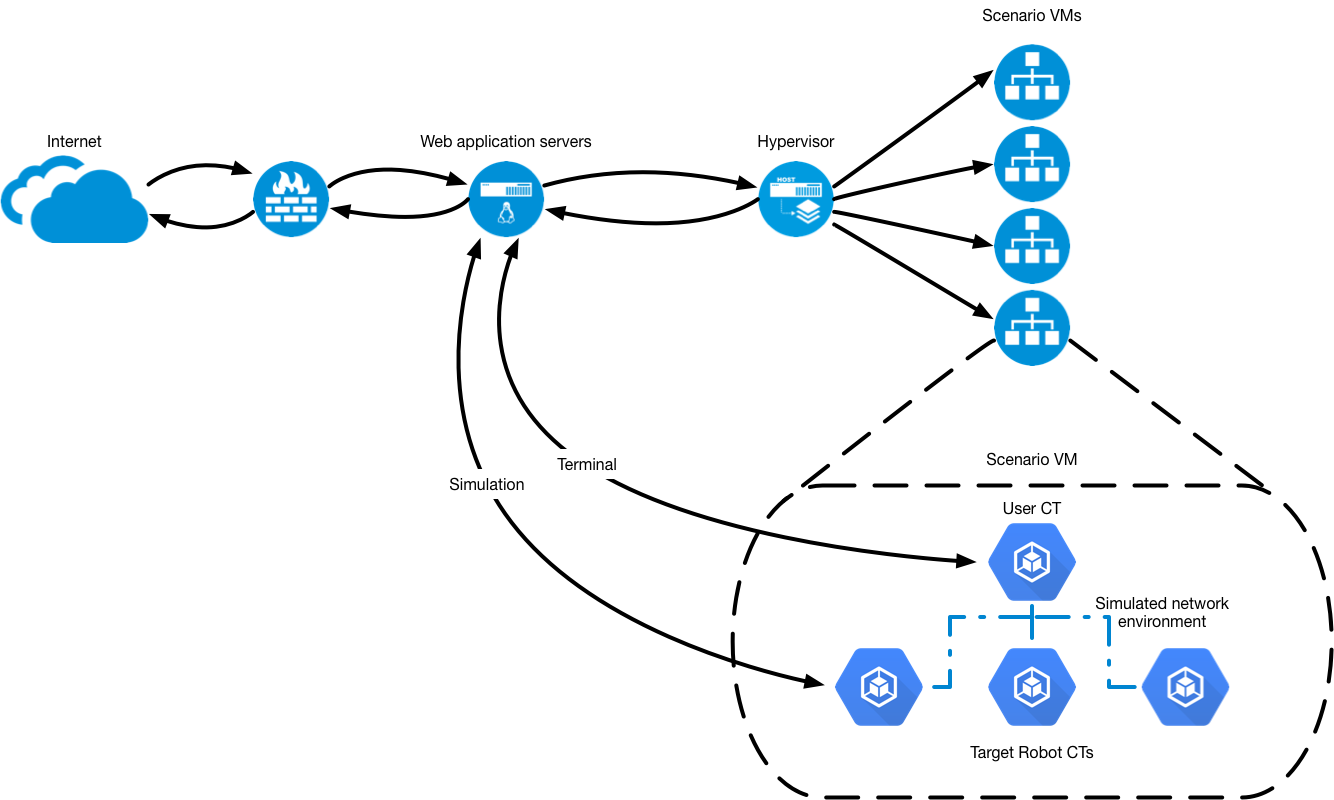}
 \caption{\footnotesize Overview of the architecture of the RCTF infrastructure and Scenarios}
 \label{fig:architecture}
\end{figure}

Several approaches were evaluated for the RCTF architecture. A trade-off between several aspects had to be considered. Mainly: speed of deployment, ability to design complex scenarios, performance of the scenario, security and containment of the infrastructure (together with the environment itself). The evaluation process  considered alternatives regarding multi-container environments and virtual machine-based environments. For the final design, a hybrid architecture between the two mentioned approaches has been adopted. RCTF uses containers for the scenario environments and virtual machines, as single units, for deployment and containment. The architecture of the Alias Robotics' RCTF is summarized in Figure \ref{fig:architecture}.\\ 

As pictured in Figure \ref{fig:architecture}, a web based application is used in conjunction with a hypervisor that orchestrates the creation of a series of scenarios for the robot hackers. The RCTF scenarios consist of self contained Virtual Machines (VMs) that host the containers that form the scenarios. This operation allows for the creation of complex scenarios, in which several ROS machines can coexist in a given simulated network environment. In order to improve the deployment speed and operational cost, linked clones based on Logical Volume Manager are being used, which allow the scenarios to be launched over certain timespans of less than a second and take little storage. \\

To avoid compromising the security and the stability of our Robotics CTF, we decided to use VMs as containers of the different scenarios. Each scenario is associated with an individual VM that communicates with the web server through two sockets, one for the terminal connection and one for the simulation (Gazebo). To facilitate the construction and reproduction of results, each scenario (VM) is built using Docker with one or several instances. In other words, our architecture comprises Virtual Machines that embed \emph{containers}. By doing this, we allow security researchers to 1) reproduce the RCTF scenarios locally and 2) study different networking setups. 

The benefits of our approach are twofold. First, by giving hackers the chance to run scenarios locally, we allow them to explore flaws without the restrictions of remote servers. Second, while the proposed scenarios aim to be as realistic as possible, there's still a gap when compared to real robotic setups. By containerizing the scenarios, we allow security researchers to deploy the challenges in a variety of networking setups. This allows to mimic real robots, further enhancing the possibilities of detecting more flaws.

Our approach also simplifies the deployment of new environments, as all the initialization is contained on each environment, without the need of a central point of orchestration, as in traditional multi-container Docker environments. From the deployment perspective, each scenario is treated as a single deployment unit, fully independent of its complexity. Moreover, the RCTF approach simplifies the migration between different orchestration and cloud platforms, and minor work is required in the management plane of the cloud provider.
After scenarios are created, the robot hacker may be provided with a terminal connection to some of the containers, in addition to, when appropriate, a connection to the Gazebo Web environment.

Alias Robotics has developed a series of RCTF scenarios, with a complexity that goes \emph{in crescendo} from requiring a very basic understanding of robot frameworks such as ROS and ROS2, to being familiar with the more complex use of a variety of hacking tools. Table \ref{rctfscenarios} provides insight about the original scenarios developed by our team. These scenarios can be tested online at \url{http://rctf.aliasrobotics.com}


\begin{table}
\caption{\footnotesize RCTF  Scenario briefing.}
\label{rctfscenarios}
\begin{tabular}{ | c | p{4cm} |p{10cm} | }
 \hline
 \textbf{Scenario} & \textbf{Weakness} & \textbf{Description} \\
 \hline
1 & Cleartext Transmission of Sensitive Information, CWE-319 & The robot hacker can play with the Robot Operating System and its publisher/subscriber architecture.\\
\hline
2 & Cleartext Transmission of Sensitive Information, CWE-319 & The whitehat can experiment with some of the ROS2.0 functionalities, and see the differences and similarities between ROS and ROS2. Especially, is noted that not setting up the security features of ROS2 leverages to the same weaknesses as in ROS1.\\
\hline
3 & Usage of weak, well known credentials & The whitehat becomes a robot-hacker and needs to prove basic knowledge of ROS and basic knowledge of the series of movies Jurassic Park to make a node publish the password somewhere.\\
\hline
4 & Undefined safety boundaries & Everyone cares about robot safety. But there is no safety without security. Gazebo visualization allows us to see "Prudencio" wearing an Alias Robotics t-shirt going too close to a hacked UR10.\\
\hline
5 & Cleartext Transmission of Sensitive Information, CWE-319 & ROS traffic is not readable from an unauthorized actor, or is it?\\
\hline
6 & CWE-78: Improper Neutralization of Special Elements used in an OS Command & A bug on the script allows performing arbitrary calls to system commands. Is it exploitable?\\
\hline
7 & \emph{coming soon!} & The robot hacker is invited to try an Alias Robotics' crafted offensive tool.\\
\hline
8 & CWE-798: Use of Hard-coded Credentials & The robot hacker could search in compiled ROS application code for hardcoded credentials used during execution.\\
\hline
9 & CWE-547: Use of Hard-coded, Security-relevant Constants
 & The hacker might be able to modify binaries in order to alter the execution and compromise the confidentiality and privacy of protected sources.\\
\hline
\end{tabular}
\end{table}

\subsection{Contributing}
In an attempt to contribute with the security community, we are open sourcing the scenarios at \url{https://github.com/aliasrobotics/rctf-list}. We envision that as new scenarios become available, the sources will remain at this repository and only a subset of them will be pushed to our web servers \url{http://rctf.aliasrobotics.com} for experimentation. We invite the community of roboticists and security researchers to play online and get a robot hacker rank.

We also invite security researchers to share their scenarios with the RCTF community, with the chance of potentially integrating them on the RCTF game. We gladly accept contributions through Pull Requests at \url{https://github.com/aliasrobotics/rctf-list}. Therein, the procedure of RCTF scenario submission is summarized, which require a short description of the goal of each scenario.

\section{Remarks and future work}
\label{conclusion}
In this work, we introduce the Robotics CTF (RCTF), a platform for robot hacking. We propose a robot hacking gamification environment, accessible from any browser, 24/7 and anywhere in the world. We describe the architecture of our platform and provide exemplary scenarios that can be customized by hackers to meet their target platforms. Moreover, we highlight that our approach allows security researchers to a) reproduce scenarios locally and b) change the networking setup to mimic their real targets.

We invite the whole security researcher community to play the RCTF and contribute with new scenarios of their own. We also warn society about the increasing relevance of robot vulnerabilities and advocate in favour of the creation of a strong robot ethical hacker community. Ultimately, we claim that robot security could benefit greatly from hacker powered security and contribute by open sourcing the existing scenarios created by our team.


\bibliographystyle{IEEEtran}
\bibliography{iclr2018_workshop}
\end{document}